\documentclass[prl,superscriptaddress,showpacs,twocolumn,floatfix]{revtex4}
\pdfoutput=1
\usepackage{times}
\usepackage[pdftex]{graphicx}
\usepackage{amsmath}
\usepackage{epsfig}
\usepackage{dcolumn}
\usepackage{amssymb}

\newcommand{\dd}{\mathrm{d}}
\def\srvo3{SrVO$_3$}

\begin{document}

\title{ 
Low-energy models for correlated materials: bandwidth renormalization from Coulombic screening
}

\begin{abstract}
We provide a prescription for constructing Hamiltonians representing the low energy physics of correlated electron materials with dynamically screened Coulomb interactions. The key feature is a  renormalization of the hopping and hybridization parameters by the processes that lead to the dynamical screening. The renormalization is shown to be non-negligible for 
various classes of correlated electron materials. The bandwidth reduction effect is necessary for connecting models to materials behavior and for making quantitative predictions for low-energy properties of solids.
\end{abstract}

\author{M. Casula}
\affiliation{CNRS and Institut de Min\'eralogie et 
de Physique des Milieux condens\'es,
Universit\'e Pierre et Marie Curie,
case 115, 4 place Jussieu, 75252, Paris cedex 05, France}
\author{Ph. Werner}
\affiliation{Department of Physics, University of Fribourg, 1700 Fribourg, Switzerland}
\author{L. Vaugier}
\affiliation{Centre de Physique Th{\'e}orique,
Ecole Polytechnique, CNRS-UMR7644, 91128 Palaiseau, France}
\affiliation{Japan Science and Technology Agency, CREST, Kawaguchi
 332-0012, Japan}
\author{F. Aryasetiawan}
\affiliation{Department of Physics, Mathematical Physics,
Lund University,
S\"olvegatan 14A,
22362 Lund, Sweden}
\author{A. Millis}
\affiliation{Department of Physics, Columbia University, 538 West, 120th Street, New York, NY 10027, USA}
\author{S. Biermann}
\affiliation{Centre de Physique Th{\'e}orique,
Ecole Polytechnique, CNRS-UMR7644, 91128 Palaiseau, France}
\affiliation{Japan Science and Technology Agency, CREST, Kawaguchi
 332-0012, Japan}


\maketitle

A key step in the theoretical analysis of strongly correlated materials is the derivation, from an
all-electron Hamiltonian 
in the continuum, of an effective model which correctly captures the physics of the  low-energy degrees of freedom.  Tremendous progress in this direction has been achieved by using  density functional theory (DFT) techniques \cite{kohn} to compute a full set of energy bands, from which a subset of correlated orbitals is abstracted for further detailed study using many-body (typically dynamical mean field (DMFT)) methods \cite{biermann-review,held-review,kotliar-review}
or LDA+U. The interaction parameters used in the many-body studies are the matrix elements of the  screened Coulomb interaction in the correlated subspace. Various methods are used to obtain the screened matrix elements, including the constrained local density approximation \cite{constrained_lda_first}, linear response \cite{constrained_lda}, or the constrained random phase approximation (cRPA) \cite{crpa}.  This DFT+cRPA+DMFT approach enables quantitative, testable theoretical predictions for correlated materials.

In this paper we show that this scheme misses an important aspect of the
physics: the downfolding produces a 
dynamically
screened Coulomb interaction which leads
to an effective model with a bandwidth that is {\em reduced} relative to the
starting (e.g. DFT) bandwidth and a low energy  spectral weight which is also
reduced. This 
effect has previously been noticed \cite{crpa,imai}. 
A similar renormalization was also discussed in the context of Holstein-Hubbard
models in Refs \cite{Takada03,Macridin04}.
We present
an explicit nonperturbative prescription for determining the renormalizations
quantitatively, and 
demonstrate that the resulting effective model provides a good description
of the low-energy part of the full (dynamically interacting) model over wide 
parameter ranges. 
Computations of the renormalizations for wide classes of
correlated electron materials indicate that their inclusion 
is crucial for a quantitative description, 
in particular resolving a long-standing discrepancy between
the cRPA estimate of the Coulomb interaction and the
values needed to describe experiments.

We first provide a demonstration for the simplest case, where the downfolding from the full band structure is to a  one-band model  with hopping amplitude $t_{ij}$ between the lattice sites  $i$ and $j$. Electrons with spin $\sigma$  in the correlated orbital localized at site $i$ are created [annihilated] by the operator $d^{\dagger}_{i \sigma}$ [$d_{i \sigma}$]. Double occupation of a given atomic site costs a Coulomb energy $U$, which is renormalized from a bare value $V$ (obtained from the site-local matrix elements of $e^2/r$ among the correlated orbitals) because of screening by degrees of freedom eliminated in the downfolding process. The interaction thus takes the general form $\frac{1}{2}\left(V \delta(\tau) +U_\textrm{ret}(\tau)\right)n_i(\tau)n_i(\tau^\prime)$. Screening is contained in the  retarded part $U_\textrm{ret}$, which is parametrized by a continuum of modes of energy $\nu$ with coupling strength $\lambda^2(\nu)=-\textrm{Im} U_\textrm{ret}(\nu)/\pi$, determined by the charge fluctuations,
\begin{equation}
\label{generalU}
U_\textrm{ret}(\tau)=-\int_0^\infty \!\!\! \dd\nu \lambda^2(\nu) 
\cosh [(\tau-\beta/2)\nu]/\sinh[\nu \beta/2].
\end{equation}
For simplicity of presentation we assume at first that there is only one important bosonic mode of energy $\omega_0$ and coupling strength $\lambda$. 
The Hamiltonian is then 
\begin{align}
\label{H}
H=& - \sum_{ij \sigma} t_{ij}
d^{\dagger}_{i \sigma} d_{j \sigma}
+ 
V \sum_i d^{\dagger}_{i \uparrow} d_{i \uparrow}
d^{\dagger}_{i \downarrow} d_{i \downarrow}
+ \mu \sum_{i \sigma} d^\dagger_{i \sigma} d_{i \sigma}
\nonumber
\\
& +
\omega_0 \sum_i b_i^{\dagger} b_i
+ \lambda \sum_{i \sigma} 
d^{\dagger}_{i \sigma} d_{i \sigma}
( b_i + b_i^{\dagger} ).
\end{align}
A Lang-Firsov (LF) transformation \cite{lang_firsov,werner_millis_hubbard_holstein} $H\rightarrow H_{LF}=e^S H e^{-S}$ with
$S = - \frac{\lambda}{\omega_0} \sum_{i \sigma} n_{i \sigma} (b_i + b_i^{\dagger})$
allows one to rewrite the model in terms of the polaron operators $c^\dagger_{i \sigma} =  \exp(\frac{\lambda}{\omega_0} (b^\dagger_i -b_i)) d^\dagger_{i \sigma}$ and $c_{i \sigma} =  \exp(\frac{\lambda}{\omega_0} (b_i - b_i^{\dagger})) d_{i \sigma}$. We note that $c$ and $c^\dagger$ obey the same fermionic anti-commutation relations as the original electronic operators ($d$ and $d^\dagger$).  Neglecting one-body terms which can be absorbed in a chemical potential shift,
we have 
\begin{equation}
\label{H_LF}
H_\textrm{LF}= - \sum_{ij \sigma} t_{ij}
c^{\dagger}_{i \sigma} c_{j \sigma}
+ 
U_0 \sum_i c^{\dagger}_{i \uparrow} c_{i \uparrow}
c^{\dagger}_{i \downarrow} c_{i \downarrow}
+ \omega_0 \sum_i b_i^{\dagger} b_i,
\end{equation}
with the screened Hubbard interaction $U_0 = V - \frac{2 \lambda^2}{\omega_0}$.

We now assert that the low energy effective model is given by {\em the projection of Eq.~(\ref{H_LF}) onto the subspace of zero-boson states},
$H_\textrm{eff}=\langle 0 | H | 0 \rangle$.  The effective model is then  
\begin{align}
\label{H_eff}
&H_\textrm{eff}
= 
- \sum_{ij \sigma} Z_B t_{ij}
d^{\dagger}_{i \sigma} d_{j \sigma}
+ 
U_0 \sum_i d^{\dagger}_{i \uparrow} d_{i \uparrow}
d^{\dagger}_{i \downarrow} d_{i \downarrow},
\end{align}
that is, an effective Hubbard model with an instantaneous  interaction corresponding to the low frequency limit of the screened interaction and a new feature, namely a  bandwidth renormalized by $Z_B = \exp (- \lambda^2/\omega_0^2)$.  An additional physical consequence of the low-energy projection is that the photoemission spectral weight in the frequency range described by the effective model is reduced by the factor $Z_B$ relative to what would naively follow from $H_{LF}$.  Mathematically, $G^\textrm{low-energy}$, the physical electron Green function in the frequency range described by the effective model, is
\begin{equation}
\label{G_eff}
G^\textrm{low-energy}_{ij}(\tau)
= - Z_B \langle T d_i(\tau) d^\dagger_j(0) \rangle_{H_\textrm{eff}},
 \end{equation}
where $ - \langle T d_i(\tau) d^\dagger_j(0) \rangle_{H_\textrm{eff}}$ is the Green's function $G^\textrm{eff}_{ij}(\tau)$ of the effective Hamiltonian $H_\textrm{eff}$ in Eq.~(\ref{H_eff}). Thus the observable spectral function $A^\textrm{low-energy}=
-\frac{1}{\pi}
\textrm{Im} G^\textrm{low-energy}(\omega-i\delta)$ becomes
\begin{equation}
\label{A_eff}
A^\textrm{low-energy}(\omega)= 
- 
\frac{Z_B}{\pi}
\text{Im} G^{\textrm{eff}}(\omega-i\delta).
\end{equation}
The physical origin is that part of the physical photoemission spectrum corresponds to the simultaneous creation of a hole and a plasmon excitation; these plasmon shakeoff processes account for the remaining  $1-Z_B$  spectral weight.

\begin{table}[b]
\caption{Critical interaction strength $U_{\textrm{crit}}^\textrm{exact}$ (presented in terms of zero frequency screened value)  needed to drive the metal insulator transition obtained from the single-site DMFT approximation to Eq.~(\ref{H}) at inverse temperature $\beta=100$ and compared to the estimate $U_{\textrm{crit}}^\textrm{eff}$ for different values of the screening frequency $\omega_0$ and strength $\lambda$. Also shown is the Lang-Firsov renormalization factor $Z_B=\exp[-\lambda^2/\omega_0^2]$. }
\label{ucrit_table}
\begin{ruledtabular}
\begin{tabular}{ d | d | d | d | d l}
\makebox[0pt][c]{$\omega_0$} & \makebox[0pt][c]{$\lambda$} & \makebox[0pt][c]{$Z_B$} & \makebox[0pt][c]{$U_{\textrm{crit}}^\textrm{exact}$ \cite{werner_millis}} & \makebox[0pt][c]{$U_{\text{crit}}^\textrm{eff}$}\\
\hline
1.5 & 0.820 & 0.74 &  2.103 &  1.891 \\
1.5 & 2.010  & 0.17 & 0.613  & 0.423 \\
2.5 & 1.330 & 0.75 &  2.085 & 1.921 \\
2.5 & 2.770 & 0.29 & 0.861 & 0.747 \\
10.0 &  3.725 & 0.87&  2.225 & 2.220 \\
10.0 &  6.465 & 0.66&  1.640 & 1.679 \\
\end{tabular}
\end{ruledtabular}
\end{table}

The effective model becomes an exact description of the low energy
physics only  when the ratio of the boson frequency $\omega_0$ to a
relevant energy $E^*$ diverges, but we find that the effective model
gives a reasonably good description even for $\omega_0/E^*$ not too
large.  As an example, Table~\ref{ucrit_table} compares exact results
(obtained using the methods of Ref.~\cite{werner_millis}) for  the
critical interaction strength $U_\text{crit}$  needed to drive a
metal-insulator transition in  single-site DMFT to the predictions of
the effective model.   In these computations we assume that the
$t_{ij}$ give a semicircular density of states with half-bandwidth
$D=1$. Combining previously computed single site DMFT results
\cite{werner_millis} with our bandwidth reduction prescription gives,
at inverse temperature $\beta=100/D$, an effective model prediction
$U^\textrm{eff}_\textrm{crit}\approx 2.55 Z_B$. 
One sees that the effective model result is within $15\%$ of the exact result except when there is strong screening and the boson  frequencies are smaller than the full bandwidth (2 in present units).

\begin{figure}[t]
\includegraphics[width=\columnwidth]{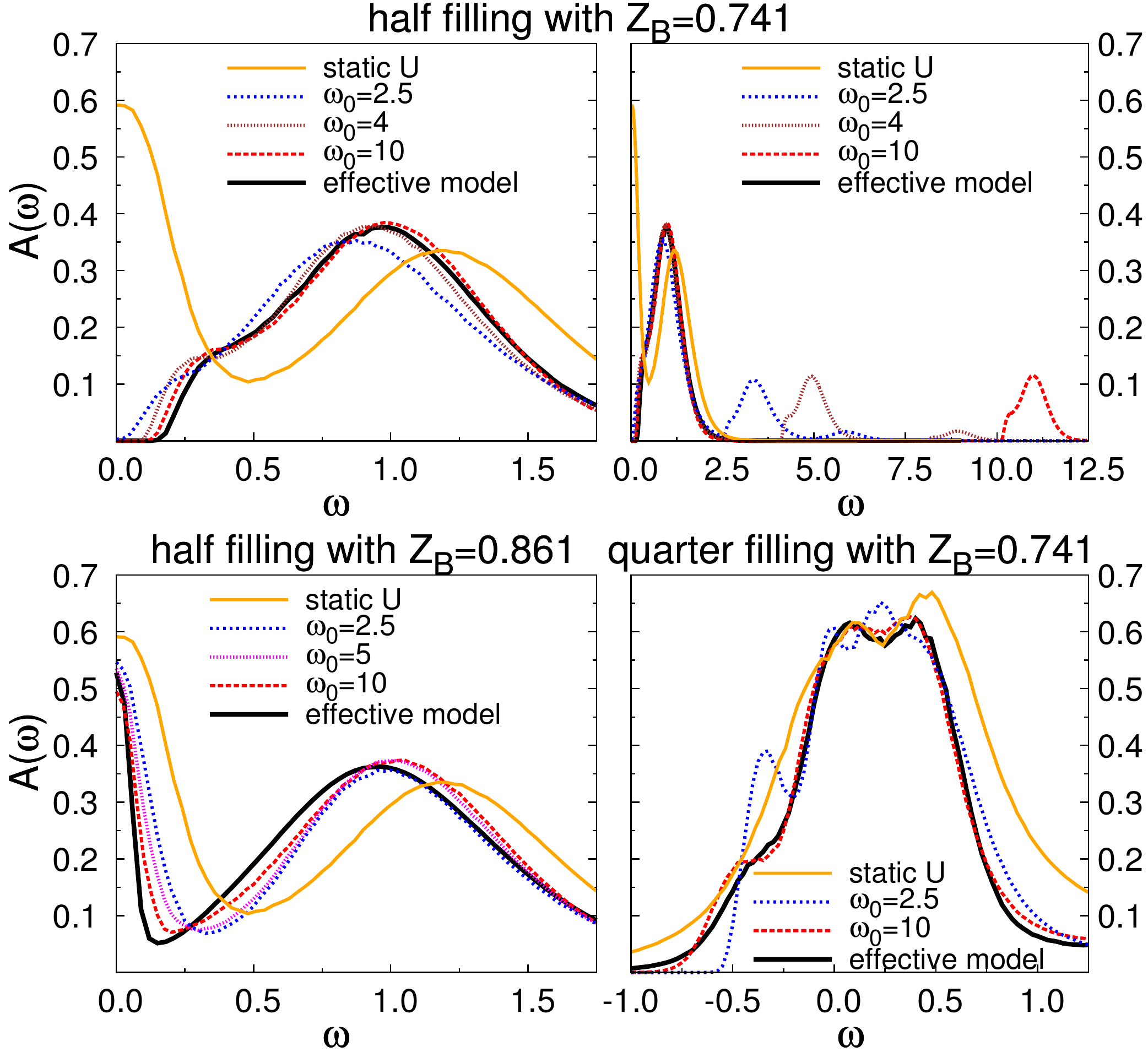}
\caption{Spectral functions computed from  Eq.~(\ref{H}) at various
  screening frequencies $\omega_0$  with $\beta=40$, screened
  interaction  $U_0=2$ and coupling constants chosen to produce the
  renormalization factor $Z_B$ as indicated.  Also shown are the
  spectral functions computed from the effective model
  (Eq.~(\ref{A_eff})) and for the static $U$ approximation.}
\label{spectrum_half}
\end{figure}

Figure~\ref{spectrum_half} compares the electron spectral function,
calculated from Eq.~(\ref{H}) with semicircular density of states
(half bandwidth $D=1$), for screened interaction $U_0=2$ with  values
of  $Z_B$ representative of typical correlated electron materials to
two approximations: the effective model defined above, and a ``static
$U$ model'' which uses the static value of the screened Coulomb
interaction but does not include the bandwidth reduction. The static
$U$ model corresponds to what is normally done in DFT+DMFT
calculations.   The analytic continuations are obtained using the
technique proposed in Ref.~\cite{casula-rubtsov-biermann}.  We see
that the effective model with bandwidth reduction $Z_B$ reproduces
very well the effective bandwidths of the Hubbard bands for all
$\omega_0$ taken into account here, which vary from $10$ down to
$2.5$. Even the smallest $\omega_0$, which is not in the antiadiabatic
regime, yields Hubbard bands qualitatively well described by the
static model with bandwidth renormalization $Z_B$. 
The static $U$ model is seen to be a poor approximation. 
\begin{table}[!ht]
\caption{Quasiparticle residue $a$ = $1/(1-\partial \textrm{Im}[\Sigma (i
  \omega)]/\partial \omega|_{\omega=0})$ computed from the effective Hamiltonian Eq.~(\ref{H_eff}) with screened $U_0=2$,  and different $\omega_0$, $Z_B$ and particle density as shown. The values in  parenthesis give the relative discrepancy $|a(\omega_0)/a(\omega_0\rightarrow \infty)-1|$. Note that the static model without bandwidth reduction (last row) is substantially incorrect. 
}
\label{Z_table}
\begin{ruledtabular}
\begin{tabular}{|l |  l | l l|} 
   \makebox[0pt][l]{}  & \multicolumn{1}{c|}{\textrm{half-filling}}  &  \multicolumn{2}{c|}{\textrm{quarter-filling}} \\
\hline
\hline
                 &  $Z_B$=0.861   &  $Z_B$=0.861 & $Z_B$=0.741   \\
\hline
\hline
$\omega_0=2.5$    &  0.137 (0.37)  &  0.635 (0.04) &
0.560 (0.10) \\
$\omega_0=3$    & 0.125 (0.32)   &  0.631 (0.03) &
0.551 (0.08) \\
$\omega_0=10$    & 0.091 (0.06)   &  0.604 (0.01) &
0.509 (0.01)\\
$\omega_0=\infty$ & 0.085  & 0.609  & 0.504\\
\textrm{static U} &      0.253           &   0.713 & 0.713 \\    
\end{tabular}
\end{ruledtabular}
\end{table}

Table~\ref{Z_table} shows the results of an alternative analysis,
carried out at the level of the quasiparticle renormalization
$a=1/(1-\partial \Sigma/\partial i \omega_n)$, which is obtained
directly from the imaginary time computations.  We see that the
``static $U$'' result gives renormalization factors in error by
factors of two or more in the half filled, strongly correlated case,
and also unacceptably large errors in the weakly correlated quarter
filled case. The effective model 
(row $\omega_0=\infty$) 
is very close to the exact result for all screening frequencies in the weakly correlated quarter-filled case and is reasonably close to the exact result even as the adiabatic limit is approached. 

Analogous arguments for a model comprising also itinerant $p$ states, and thus hopping parameters ${\cal T}_{pp}$, ${\cal T}_{pd}$ and ${\cal T}_{dd}$ lead to a renormalization of each $d$ operator by the factor $\sqrt{Z_B}=\langle 0 | \exp(\frac{\lambda}{\omega_0} (b_i - b_i^{\dagger})) | 0 \rangle$ so that the hopping part of the one-particle Hamiltonian is renormalized as
\begin{equation}
\left( p^\dagger d^\dagger \right) \left( \begin{array}{cc}
{\cal T}_{pp}  & \sqrt{Z_B} {\cal T}_{pd} \\
\sqrt{Z_B} {\cal T}_{pd}^\dagger & Z_B {\cal T}_{dd} \end{array} \right) 
\left( \begin{array}{c} 
p \\
d \end{array} \right),
\label{matrix_transformation}
\end{equation}
where the site dependence of each orbital species is  not explicitly denoted. Equation~(\ref{matrix_transformation}) shows that the bandwidth reduction implied by our effective model  happens in a non trivial way in the case of the multi-band models usually dealt with in first-principles calculations. 

The arguments we have given are readily generalized to the case of an arbitrary dynamical interaction. The representation of Eq.~(\ref{generalU}) corresponds to a continuum of boson excitations, $b_i(\nu)$, one for each frequency $\nu$ in the screening process, with 
coupling $\lambda(\nu)$.  We  then apply a generalized LF transformation  obtaining 
\begin{eqnarray}
U_0 & = & V + 2/\pi \int_0^\infty \!\!\! \dd\nu ~ \textrm{Im}
U_\textrm{ret}(\nu) /\nu, 
\label{U_0}\\
Z_B & = & \exp \left( 1/\pi  \int_0^\infty \!\!\! \dd\nu
  ~  \textrm{Im} U_\textrm{ret}(\nu) /\nu^2 \right).
\label{Z_B}
\end{eqnarray}
Matching this to the single mode formula implies a characteristic frequency
\begin{eqnarray}
\omega_0&=& \frac{\int_0^\infty \!\!\! \dd\nu
  ~ \nu \textrm{Im} U_\textrm{ret}(\nu) /\nu^2}
  {\int_0^\infty \dd\nu  ~ \textrm{Im} U_\textrm{ret}(\nu) /\nu^2}.
  \label{omega0}
\end{eqnarray}

\begin{table}
\caption{\label{tab3}
Boson renormalisation factor $Z_B$, characteristic frequency
$\omega_0$ [eV], 
bare interaction $V$ [eV],
zero-frequency screened interaction $U_0$ [eV] as calculated within
the cRPA, in the
implementation of Ref. \cite{loig}.
For the oxide and sulfide compounds (except SrMnO$_{3}$), 
data refer to a model comprising
only the t$_{2g}$ states, where $U$ is defined as the average over the diagonal entries of the Hubbard
interaction matrix $U_{mmmm}$. For the pnictide compounds, as well as
for SrMnO$_{3}$ and CuO, a hybrid ``d-dp''
model in the notation of Ref. \cite{aichhorn-09,loig} was constructed and $U (=F_0)$ is defined as the average over all density-density
interaction matrix elements.
Experimental lattice structures (rutile in the case of VO$_2$, hexagonal
lattice in the case of TaS$_2$) were used except in the cases of Sr$_{2}$VO$_{4}$, LaVO$_3$
and SrMnO$_3$, where an undistorted (double) perovskite structure was adopted.
The column headed $U_\textrm{lit}$ gives $U$ values obtained via a 
variety of methods other than cRPA claimed in the literature to give
quantitative agreement with experiment when used in 
DFT+DMFT (oxides, sulfides and pnictides) 
or
DFT+U calculations (SrMnO$_{3}$ and CuO)  within
the same correlated subspace, but
without the band renormalization physics.
}
\begin{ruledtabular}
\begin{tabular}{|l|c c c c r l|}
      &  $Z_B$ & $\omega_{0}$ &
$V$ & $U_{0}$ & $U_{\textrm{lit}}$ &\\
\hline
SrVO$_{3}$              & 0.70  &  18.0    &   16.5  & 3.3  & 4 - 5 & \cite{liebsch, lechermann-d1,pavarini-d1, sekiyama} \\
Sr$_{2}$VO$_{4}$      & 0.70 & 18.1 & 15.7 & 3.1 & 4.2 & \cite{arita07} \\
LaVO$_{3}$              & 0.57 & 10.3 & 13.3  & 1.9 & 5 & \cite{pavarini-d2}\\
VO$_{2}$                  & 0.67 & 15.6 & 15.2 & 2.7 & 4 & \cite{biermann_vo2,tomczak} \\
TaS$_{2}$                & 0.79 & 14.7 & 8.4 & 1.5 &  & \\
SrMnO$_{3}$           & 0.50  & 13.3  &  21.6 & 3.1 &   2.7 &  \cite{lee-rabe10}  \\
BaFe$_{2}$As$_{2}$  &  0.59   & 15.7 &  19.7  & 2.8 &  5 & \cite{yin-kotliar-NatMat11}   \\
LaOFeAs                 &   0.61  & 16.5  & 19.1 & 2.7  & 3.5 - 5 & \cite{haule08,anisimov09,yin-kotliar-NatMat11}   \\
FeSe                        &  0.63 & 17.4 &   20.7 & 4.2  &  4 - 5 & \cite{craco08,yin-kotliar-NatMat11} \\
CuO                        & 0.63 & 21.1 & 26.1 & 6.8 & 7.5 & \cite{wu06} 
\end{tabular}
\end{ruledtabular}
\end{table}

Our theory has important implications for electronic structure calculations for correlated materials.  
Table (\ref{tab3}) presents our results for  
$\omega_0, Z_B$ and $U$ values for a range of compounds
calculated using the cRPA method \cite{crpa}, in the implementation 
of Ref.~\onlinecite{loig}.
Typical $Z_B$ values for oxides or pnictides lie in the range of $\sim 0.6-0.7$ indicating substantial renormalization of the low energy  bandwidths relative to DFT calculations, even though the screening frequencies $\omega_0$ are typically high. Standard DFT+DMFT calculations are available for most of the compounds. As shown in Table~\ref{tab3}, in these calculations, obtaining agreement with experimental results for mass enhancements and metal-insulator phase diagrams has required the use of $U$ values substantially ($\sim 40 \% $) larger than the low-frequency Hubbard interactions calculated from cRPA. For example, for SrVO$_3$, LDA+DMFT  calculations with $U$ ranging from 4 to 5 eV were found to yield  good agreement with experiments \cite{liebsch, sekiyama, pavarini-d1} (instead of the cRPA value of 3.5 eV).  
Similarly, in VO$_2$, $U=4.0$ eV was used \cite{biermann_vo2,tomczak}  instead of $U=2.7$ eV.
We believe that the difference arises because the previous literature did not incorporate the bandwidth reduction effect, and artificially compensated this by increasing $U$. 
The one apparent exception is SrMnO$_3$, where the $U$ value
quoted in Ref.~\onlinecite{lee-rabe10} was chosen to be consistent with the
magnetic moment but gaps or other dynamical properties were not
studied. A more recent work of a t$_{2g}$-only model required
a rather larger value of 3.5 eV, but overlap with $e_g$ bands
precludes a cRPA estimate of $Z_B$ in this case.

\begin{figure*}[ht]
\includegraphics[width=2\columnwidth]{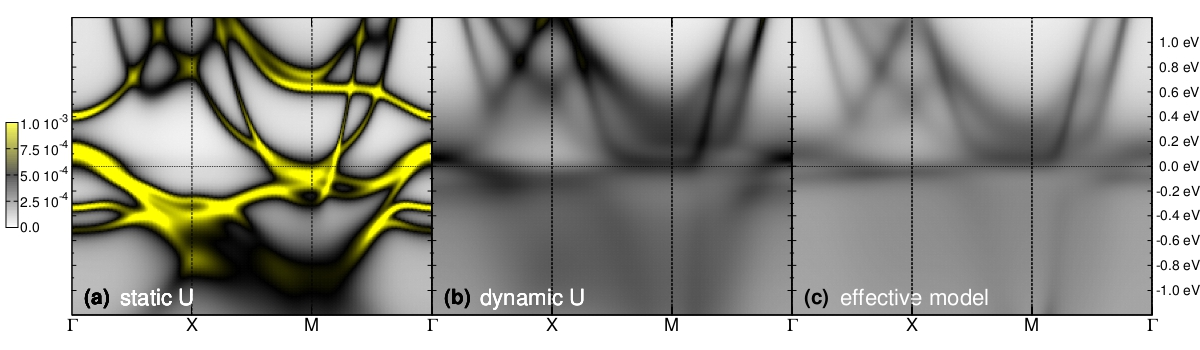}
\caption{$k$-resolved  
spectral function for K$_x$Ba$_{(1-x)}$Fe$_2$As$_2$ at
optimal doping $x=0.4$ and $\beta=20$eV$^{-1}$, reported for a 
static $U$ standard DMFT calculation (panel (a)),
the DMFT calculation with dynamic $U(\omega)$ (panel (b)), 
and the DMFT calculation for our effective low-energy model.
In all calculations, 
the static limit of $U(=F_0)$ is $U(0)=2.84$ eV, and $J=0.68$ eV.
In the effective model, the double counting correction is
set to match the $d$-electron number of the dynamical calculation. 
}
\label{kpath_122}
\end{figure*}

Figure~\ref{kpath_122} shows another illustration of the bandwidth renormalization phenomenon, comparing the spectral function of optimally doped BaFe$_2$As$_2$ obtained with the ``static $U$" approximation (panel (a)) to the full treatment of the dynamic $U$, as explained  in Ref.~\onlinecite{our_nature} (panel (b)), and the effective  model (panel (c)). Comparison of panels (a) and (b) shows that screening has a substantial effect on the band structure, shifting the energy positions of  bands and band crossings to a significant extent. (The model with screening also has an increased broadening resulting from a change in proximity to a spin freezing line whose position depends very sensitively on parameters  \cite{our_nature}; this effect is not of primary interest here). Comparison of panels (b) and (c) shows that the effective model captures the changes in band energies very well, and also reproduces the change in lifetime. 

To summarize, in this Letter we showed that the low energy effective
Hamiltonian relevant to correlated electron materials involves two
renormalizations: a reduction, to a value smaller than the isolated
atom value, of the on-site Coulomb interaction and a reduction, to a
value smaller than the band theory value, of the bandwidth. The
reduction of the onsite Coulomb interaction is a straightforward
consequence of screening by high energy degrees of freedom and has
been discussed in many works. The reduction of the bandwidth  is a
more subtle effect, which has  important consequences for the low
energy physics, including a reduction in the amplitude, and a
narrowing of the width of the low-energy part of the electron spectral
function, as well as a shift in the location of the metal-insulator
transition. We have provided a precise prescription for obtaining the
bandwidth reduction and have tested our low-energy effective
description against numerically exact dynamical mean field solutions
of Hubbard models with full dynamic $U$ in a range of parameters
relevant for correlated materials. Important open questions are the
issues of full charge-self consistency and the related double counting correction, both of which require knowledge of  physics at energy scales above the range of validity of the low-energy effective model. This is the subject of current research.

We acknowledge useful discussions with T. Miyake, H. Jiang and G. Sawatzky. This work was supported by the French ANR under project SURMOTT, IDRIS/GENCI under Grants 2012096493
and 201201393, SNF Grant PP0022-118866, FOR 1346 and the US 
DOE under grant BES ER-046169.

\end{document}